\documentclass[conference]{IEEEtran}
\usepackage{graphicx}
\usepackage{amsmath}
\usepackage{amsfonts}
\usepackage{bm}
\usepackage{graphicx}
\usepackage{slashbox}
\usepackage{threeparttable}
\usepackage{color}
\usepackage{multirow}
\usepackage{pifont}
\usepackage{tabularx}
\usepackage[square, comma, sort&compress, numbers]{natbib}
\usepackage[caption=false]{subfig}
\usepackage{color}
\usepackage{comment}

\ifCLASSINFOpdf
\else
\fi
\begin{document}
%
\title{Towards a Multi-array Architecture
	for Accelerating Large-scale Matrix
	Multiplication on FPGAs}

\author{\IEEEauthorblockN{Junzhong Shen$^{1,2}$, Yuran Qiao$^{1,2}$, You Huang$^{1,2}$, Mei Wen$^{1,2}$ and Chunyuan Zhang$^{1,2}$}
\IEEEauthorblockA{$^{1}$College of Computer, $^{2}$National Key Laboratory for Parallel and Distributed Processing\\
National University of Defense Technology,
Changsha, China 410073\\
Email: shenjunzhong@nudt.edu.cn}
}


%


\maketitle

\begin{abstract}
Large-scale floating-point matrix multiplication is a fundamental kernel in many scientific and engineering applications. Most existing work only focus on accelerating matrix multiplication on FPGA by adopting a linear systolic array.
This paper towards the extension of this architecture by proposing a scalable and highly configurable multi-array architecture. In addition, we propose a work-stealing scheme to ensure the equality in the workload partition among multiple linear arrays. Furthermore, an analytical model is developed to determine the optimal design parameters.
Experiments on a real-life convolutional neural network (CNN) show that we can obtain the optimal extension of the linear array architecture.


\end{abstract}


%
\IEEEpeerreviewmaketitle

\section{Introduction}
	Large-scale floating-point matrix multiplication is widely used in many complicated computation tasks such as scientific computing \cite{govindu2005library} and deep learning \cite{lei2016}.
	Recently, field-programmable gate arrays (FPGAs) have become a particularly attractive option for accelerating large-scale matrix multiplication due to their reconfigurability and abundant logic resources.
    Previous studies \cite{Jang2002,Zhuo2004,Jang2005,Dou2005,Kumar2009,Jovanovic2012,liu2017throughput} have primarily focused on accelerating matrix multiplication on FPGA by using an efficient architecture, i.e. the one-dimensional systolic array. This architecture was demonstrated successfully in matrix multiplication acceleration, which contributes to low bandwidth requirement and good scalability of the accelerator designs.
	
    In this paper, we focus on the extension of the linear array architecture. According to our studies, there exist three approaches to extend the linear array architecture: 1. increasing the length of the linear array; 2. adopting multiple parallel linear arrays; 3. combining approaches 1 and 2.  
   	As a result, the design space is significantly expanded, which make it more difficult to reach the optimal solution than previous work. In addition, the computation efficiency of the accelerator is hard to be ensured if we adopt a fixed architecture for various problem sizes. 
	This paper addresses these challenges by proposing a highly configurable and scalable multi-array architecture,
	which allows users to dynamically change the number of PEs in used as well as the number of parallel PE arrays. In addition, a work-stealing scheme is adopted to guarantee the equality of the workloads partition among the PE arrays. To determine the optimal design option for the proposed architecture, we also propose an analytical model to evaluate the realistic memory bandwidth as well as quantifying data traffic volumes. 

	The rest of this paper is organized as follows. We review the background in Section II. The proposed multi-array architecture is illustrated in Section III. Section IV presents the performance model. The experimental results are presented in Section V. Section VI concludes this paper.

	\begin{figure*}
		\centering
		\includegraphics[width=0.8\textwidth]{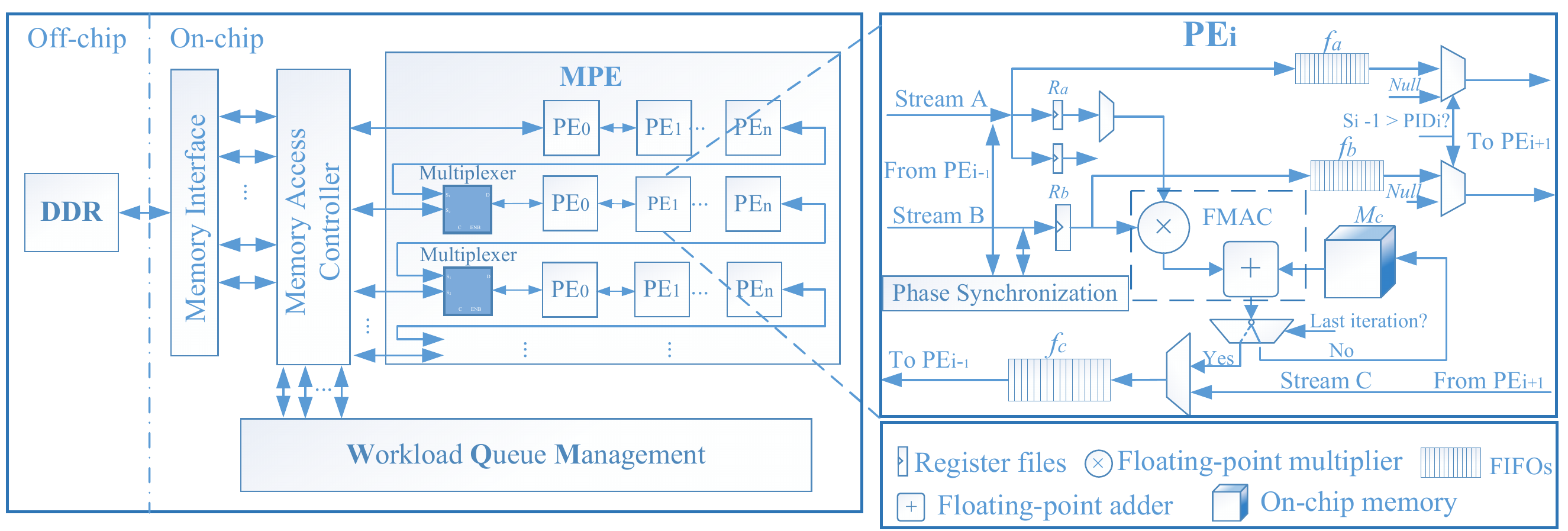}
		\caption{An overview of our proposed architecture.}
		\label{Proposed architecture}
		\vspace{-1.4em}
	\end{figure*}

	\section{Background}
	
	In this paper, we focus on the dense matrix multiplication (MM) algorithm: $\textbf{C}=\textbf{A} \times \textbf{B}$, where matrix $\textbf{A}$ $\in \mathbb{R}^{M \times K}$, $\textbf{B}$ $\in \mathbb{R}^{K \times N}$ and \textbf{C} $\in \mathbb{R}^{M \times N}$, respectively. 
	Equation 1 describes the computation pattern of this algorithm:
	\begin{equation}\label{key}
		\setlength{\abovedisplayskip}{3pt}
		\setlength{\belowdisplayskip}{3pt}
	c_{i,j}=\sum_{k=1}^{n}a_{i,k}\cdot b_{k,j}. 
	\end{equation}
	
	Since it is infeasible to store the entire input and output matrices on the FPGA, abundant researches \cite{Dou2005,Jang2005,Zhuo2004,Zhuo2007,Kumar2009,Wu2010b,Jovanovic2012} adopt a similar block matrix multiplication algorithm in their designs. Here, we introduce the block matrix multiplication algorithm in Dou \cite{Dou2005}, which has been proved to be successful in matrix multiplication acceleration.
	Initially, matrix $\textbf{A}$ is split into $\lceil M/S_i \rceil$ sub-blocks (namely $SA$) of size $S_i \times K$ and matrix \textbf{B} is partitioned into $\lceil N/S_j \rceil$ sub-blocks (namely $SB$) of size $ K \times S_j $. In this way, the result matrix $\textbf{C}$ can be calculated by performing sub-block matrix multiplications on the $SA_i$ and $SB_j$, where $i \in [1, \lceil M/S_i \rceil]$ and $j \in [1,\lceil N/S_j \rceil]$.
    The basic idea of this algorithm is to split the multiplication of $SA_i$ and $SB_j$ into multiple inner-product operations of two vectors, i.e $U_k$ and $V_k$, where $U_k$ is the $k^{th}$ column of $SA_i$ and $V_k$ is the $k^{th}$ row of $SB_j$ ($k \in [1, K]$).   
    Here we define $C_{i,j}$ as the product of $SA_i$ and $SB_j$, 
    then $C_{i,j}$ can be calculated by accumulating $C_1$, $C_2$,.., and $C_K$ iteratively, shown as  
    \begin{equation}
    		\setlength{\abovedisplayskip}{3pt}
    		\setlength{\belowdisplayskip}{3pt}
         C_{i,j} = SA_i \times SB_j = \sum_{k=1}^{K}V_k \times U_k 
    \end{equation}

	\section{Proposed Architecture}
	\label{section architecture}
	
	Fig.~\ref{Proposed architecture} presents an overview of our proposed architecture for acceleration of matrix multiplication.
	Due to limited amount of on-chip memory on FPGAs, the source data and final results are stored in the off-chip memory (i.e. the DDR). 
    It can be seen that the accelerator is composed of several modules, namely the Memory Access Controller \textit{(MAC}), the Workload Queue Management (\textit{WQM}), and the Matrices Processing Engine (\textit{MPE}). 
	
	\subsection{MPE Design}
	
	The MPE is the kernel computation module of our architecture. As shown in Fig.~\ref{Proposed architecture}, the \textit{MPE} module consists of several linear arrays of PEs, in addition with some multiplexers placed between adjacent PE arrays.
	
	We apply two operation modes in the two adjacent PE arrays, namely the \textit{Independent} mode and the \textit{Cooperation} mode.
	In the \textit{Independent} mode, the multiplexer between the PE arrays is disabled, meaning that the PE arrays can execute computation tasks independently without any data communication.
	While in the \textit{Cooperation} mode, the multiplexer between the PE arrays is enabled. As a result, the data paths of the PE arrays are connected by the multiplexer. As shown in Fig.~\ref{Proposed architecture}, the PE array that placed behind a multiplier can fetch data from the proceeding PE array in this mode.
    In \textit{Cooperation} mode, the required memory bandwidth of the PE arrays is lower since the PE arrays share the same memory interface when they are connected. 
	In addition, larger block sizes can be supported in the \textit{Cooperation} mode since the number of PEs in the connected array has increased. 
	Note that the multipliers are initialized by the host CPU, meaning that the multi-array architecture is highly configurable. More importantly, our architecture preserves the scalability of the linear array architecture.
		
	The fully pipelined structure of PE is presented in the right part of Fig.~\ref{Proposed architecture}. It can be seen that the PE consists of two sets of data registers for input data buffering, three First-In-First-Outs (FIFOs) for delivering data between PEs, local memory for temperate data storing, and floating-point multiply-and-accumulate unit ($FMAC$). Different from previous studies, we implement additional control units to support arbitrary block size. In addition, we implement a phase synchronization unit ($PSU$) to guarantee the correctness of the final results when the block sizes for matrices $\textbf{A}$ and $\textbf{B}$ are different. By conditionally inserting stalls into the computation pipeline of the PE, the $PSU$ ensures that the $k^{th}$ column of $SA$ and $k^{th}$ row of $SB$ are fetched into each PE simultaneously.
	The dataflow in each PE consists of three stages: 
	
	\textbf{Prefetch}. In this stage, the PE picks up the corresponding element in $V_1$ (i.e. the first column of $SA$) based on the PE identifier (PID), then stores the data into the local register $R_a$. For instance, $PE_1$ with $PID = 1$ will picks up the second element in $V_1$.  
	
	\textbf{Compute}. In this stage, the $k^{th}$ row of $SB$ (i.e. $U_k$) and the $(k+1)^{th}$ column of $SA$ (i.e. $V_{k+1}$) are fetched into the PE simultaneously, where $1 < k \le K$.
	The buffered element in $R_a$ is multiplied with all the elements of $SB_k$ in order. Therefore, the data buffered in $R_a$ is reused $S_j$ times. In the meantime, the PE also buffers the corresponding data in $V_{k+1}$ into $R_a$. Note that we apply double buffering in $R_a$ to overlap buffering data of the next iteration and computation of the current iteration.
	The products of the multipliers in \textit{FMAC} are then added with the intermediate results generated in the previous iteration, which are stored in the local memory $M_c$.
    Finally, the newly sums are written back into the memory $M_c$.
    Note that in the last iteration (i.e. $k=K$), the final results are written into FIFO $f_c$ instead of memory $M_c$.
	
	\textbf{Write back}. 
	In this stage, the PE sends its local results to the proceeding PE or the \textit{MAC} module (only $PE_0$) from the $f_c$. As a results, the result data are delivered from the end of each independent PE arrays to the \textit{MAC} module.
		
	\subsection{WQM Design}
	The \textit{WQM} module is responsible for workloads assignment for the PE arrays.
	It manages multiple workload queues to buffer the computation tasks for the PE arrays. Note that one workload queue corresponds to one PE array.
	For the proposed multi-array architecture, the steadiness of an even partition of workloads among PE arrays is the key to achieve better performance. 
However, it is difficult to ensure that the workloads are always equally partitioned during the whole computation procedures of PE arrays. For example, sometime the PE array which is assigned with less workloads could obtain higher memory bandwidth, which may worsen the inequality of workloads partition. As a result, the system performance would bottlenecked by the PE array with the most workloads. To address this issue, we adopt the work-stealing scheme  \cite{blumofe1999scheduling} in the design of the \textit{WQM} module.

	The basic idea of the work-stealing scheme is to enable an idle PE array to acquire computation tasks from the overloaded PE array(s). 
	\begin{figure}[!t]
		\centering
		\includegraphics[width=3.0in]{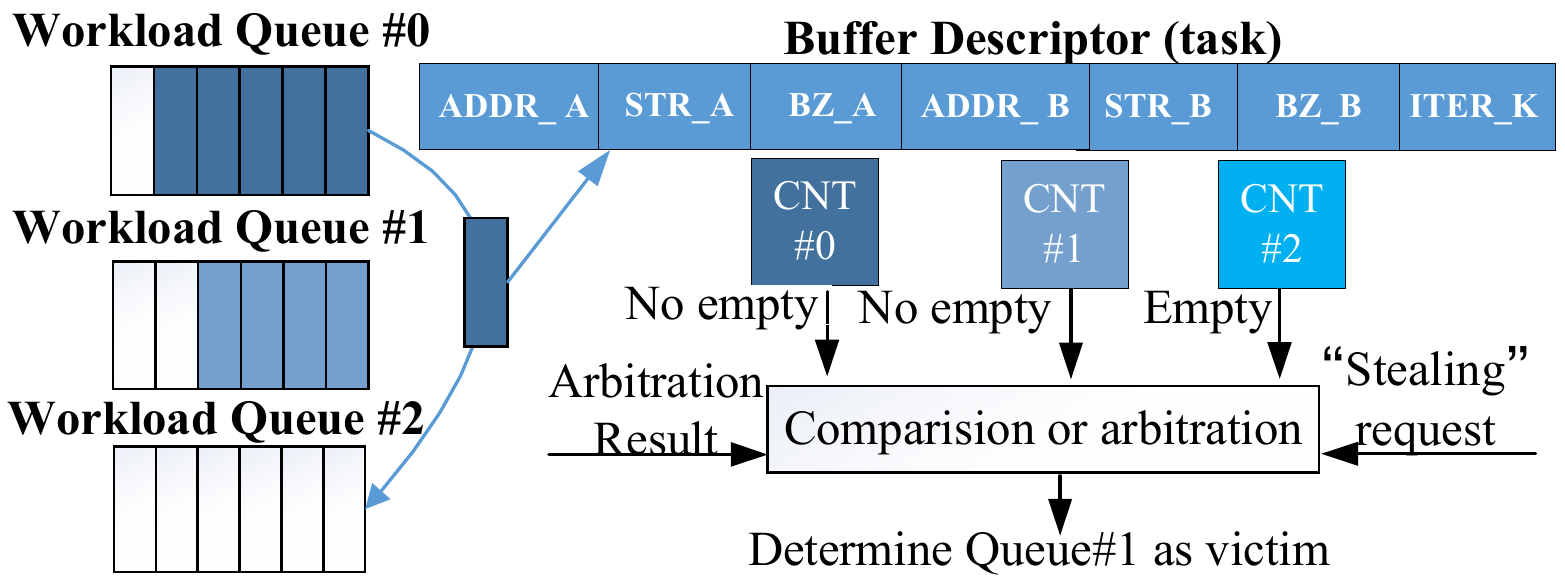}
		\caption{Illustration of our proposed work-stealing scheme.}
		\label{Work-stealing}
		\vspace{-1.4em}
	\end{figure}
	Fig.~\ref{Work-stealing} depicts the working procedure of the work-stealing scheme.
    Note that a controller (omitted in Fig.~\ref{Work-stealing}) is designed to manage the workload delivery among the workload queues.
	It can be seen that a counter is implemented to record the number of tasks for each workload queue. Once the controller detects that a workload queue becomes empty, it will immediately steal a task from a non-empty queue, then load it into the empty workload queue. 
	If there exist more than one non-empty queues, the controller will select the workload queue that with the most workloads as target, by comparing the corresponding counters of the workload queues. It is important to note that we implement a round-robin arbiter in the controller to arbitrate multiple concurrent work-stealing requests. 
	The controller repeats the detection and arbitration during the entire computation procedure of the PE arrays. 
	
	\subsection{MAC Design}
	The \textit{MAC} module is responsible for managing data transfer between the external memory and the accelerator.
    As shown in Fig.~\ref{Work-stealing}, the workloads executed by the \textit{MAC} module are organized by a self-defined data structure named \textit{buffer descriptor}.
	A \textit{buffer descriptor} contains the following parameters: $ADDR$ specifies the memory locations that store the sub-matrices; $STR$ specifies the stride of each memory transfer; $BZ$ specifies the block sizes and $ITER\_K$ specifies the iteration ($K$). 
	
	As mentioned in the above context, elements of matrix $\textbf{A}$ are fetched into the PE arrays in column-major order. However, the matrix $\textbf{A}$ is stored in row-major order. Therefore, the access of matrix $\textbf{A}$ may cause inefficient memory bandwidth utilization. 
	To improve the effective memory bandwidth, we transpose matrix $\textbf{A}$ to allow its data to be fetched in row-major order.
	In this way, 
    the burst transfer mode that favored by the external memory can be used to access both matrices $\textbf{A}$ and $\textbf{B}$.
	As a result, the memory bandwidth for the accelerator is significantly improved, which contributes to performance improvement of the overall system.
	
	\section{Performance Modeling}
	In this section, we will illustrate	how to determine the optimal solution of mapping the block matrix multiplication algorithm onto the multi-array architecture.
		
	Let the bandwidth of the off-chip memory be $BW$ (MB/s), the number of PE in a single PE array be $P$ (when all the multiplexers are disabled), the number of PE arrays work in parallel be $N_{p}$, block size of the matrix $\textbf{A}$ (on rows) be $S_{i}$ and block size of the matrix $\textbf{B}$ (on columns) be $S_{j}$. 
	For $\textbf{A}$ of size $M \times K$ and $\textbf{B}$ of size $K \times N$, 
	the average number of sub-block matrix multiplications performed by each PE array can be expressed as
	\begin{equation}
	 N_{work} = \lceil \frac{1}{N_p} \times \lceil \frac{M}{S_i} \rceil \times \lceil \frac{N}{S_j} \rceil \rceil. 
	\end{equation}
	Note that we pad matrices $\textbf{A}$ and $\textbf{B}$ with zeros if $M$ and $N$ are not integer multiples of $S_i$ and $S_j$. 
	In addition, the time (in seconds) taken to load a workload (i.e. $SA_i$ and $SB_j$) and write back the corresponding $C_{i,j}$ can be calculated by
	\begin{equation}
	 T_{work} = \frac{4 \times (S_i \times K + S_j \times K + S_i \times S_j)}{BW}. 
	\end{equation}
	To simplify the model, we assume that all the workloads are equally partitioned. Therefore, the time taken to transfer data between the external memory and the PE arrays can be expressed as
	\begin{equation}
	T_{trans} = N_{work} \times T_{work}. 
	\end{equation}
	According to the data path described in section III, the computation time $T_{compute}$ (in seconds) of a single PE array can be determined as follows:
	\begin{equation}
	T_{compute} = \frac{N_{work} \times (S_i + max\{S_i,S_j\} \times K + Stage_{fmac})}{F_{acc}},
	\end{equation}
	where $Stage_{fmac}$ denotes the stages of the computation pipeline in each PE, and $F_{acc}$ is the working frequency of the accelerator.
	Since the memory access and computation process  are overlapped in our architecture, it is difficult to directly estimate the execution time of the accelerator. However, the lower bound and upper bound of the execution time $T_{total}$ can be determined by
	\begin{equation}
	 T_{compute} < T_{total} < T_{trans} + T_{compute}. 
	\end{equation}
	To simplify the discussion on the parameters that affect $T_{total}$, we assume $S_i = S_j$ for the rest of this paper.
   It can be inferred that the attainable memory bandwidth $BW$ for each PE array is mainly affected by $N_p$ and $S_i$, which can be expressed by
	\begin{equation}
	 BW = f(N_p, S_i).
	 \end{equation}
     This is because $S_i$ determines te burst length of memory access, and $N_p$ affects the conflicts of memory accesses of the PE arrays. 
	 From equations 3, 4, 5, 6 and 8, it can be seen that $N_p$ and $S_i$ are the key factors that affect the performance of our accelerator.   
To reduce the size of design space, we consider the constraints on $N_p$ and $S_i$.
	We observe that there exists a relationship between $S_i$ and $N_p$. 
	For better understanding, we denote $P_m=4$ as the maximum number of the independent PE arrays (when all the multiplexers are disabled), then the relationship between $N_p$ and $S_i$ can be determined as
	\begin{equation}
	\left\{
	\begin{array}{ll}
	N_p \in \{1,2,3,4\}, & if \quad 1 \le S_i \le P \\
	N_p \in \{1,2\},     & if \quad P < S_i \le 2P\\
	N_p = 1,             & if \quad 2P < S_i \le 4P \\
	\end{array}. \right. \label{range} 
	\end{equation}
	To this end, the size of design space can be reduced with the assistance of equation \ref{range}.
	Given the fixed problem size and $P_m*P$ (i.e. the total number of PEs),
	the proposed analytical model can be used to determine the optimal $<S_i, N_p>$ that minimizes the range of $T_{total}$.
	
	\section{Experimental Results}
	The FPGA platform used in our experiment is the Xilinx VC709 board, which equips with a XC7VX690T FPGA and two DDR3 DRAMs. In addition, all synthesized results are obtained from Xilinx Vivado	2016.4.	
		\begin{table}[!t]
    		\caption{Post-synthesis resource utilization.}
			\label{resource}
			\begin{tabular}{|l|c|c|c|c|}
				\hline
				\parbox[l]{7mm}{\textbf{Resource}} & \textbf{DSP48Es} & \textbf{BRAMs} &   \textbf{Flip-Flops} & \textbf{LUTs} \\ \hline
				Utilization   & 1032 & 560.50  & 292016 & 192493 \\ \hline
				percentage(\%)  & 28.67 & 38.13 &  33.70 & 44.44 \\ \hline
			\end{tabular}
		\end{table}
	
\begin{figure}[!t]
		\centering
		\includegraphics[width=3.0in]{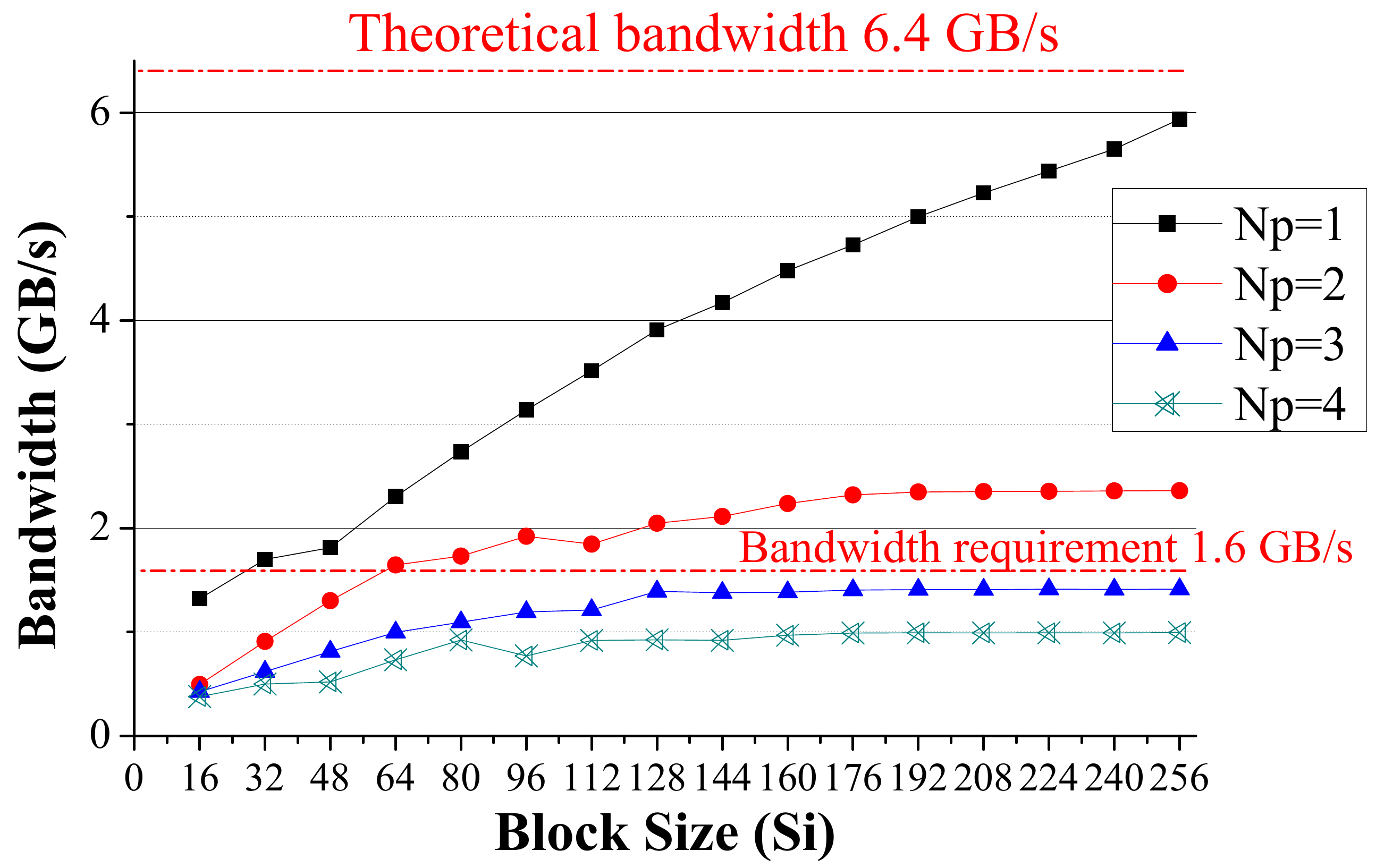}
		\vspace{-0.5em}
		\caption{Effective memory bandwidth.}
		\label{Bandwidth}
		\vspace{-1em}
	\end{figure}

		
	In order to quantify function $f$, we evaluate the average effective memory bandwidth of a PE array in terms of block sizes and number of PE arrays. 
	 As shown in Fig.~\ref{Bandwidth}, two observations can be found. First,  the effective memory bandwidth goes up with the increase of block size. Second, the effective bandwidth declines when we increase the number of PE arrays. 
	
	\begin{figure}[!t] 
		\centering
		\vspace{-1em}
		\subfloat[$N_p=1$]{\includegraphics[width=.245\textwidth]{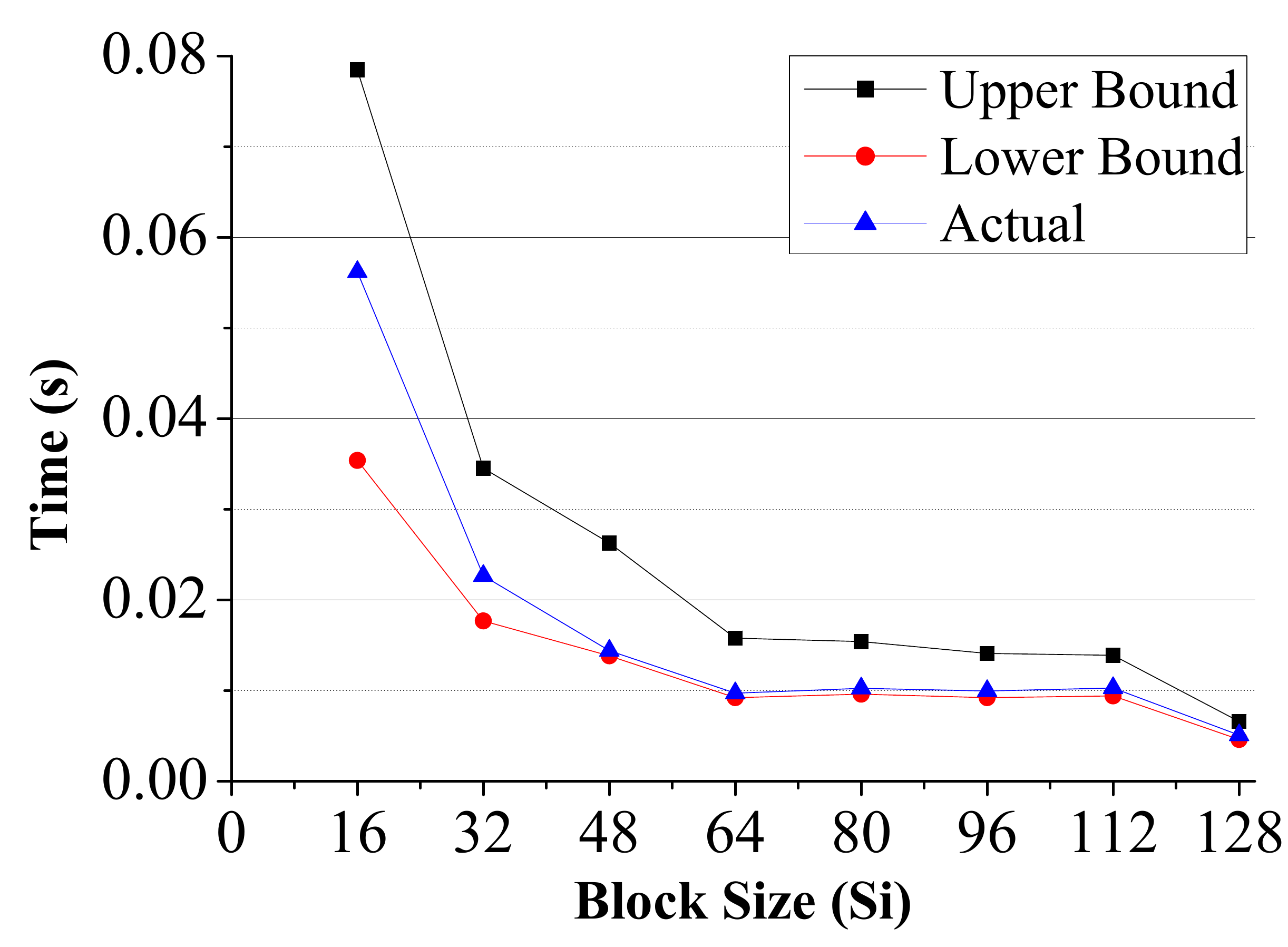}
			\label{256x1}}	
		\subfloat[$N_p=2$]{\includegraphics[width=.245\textwidth]{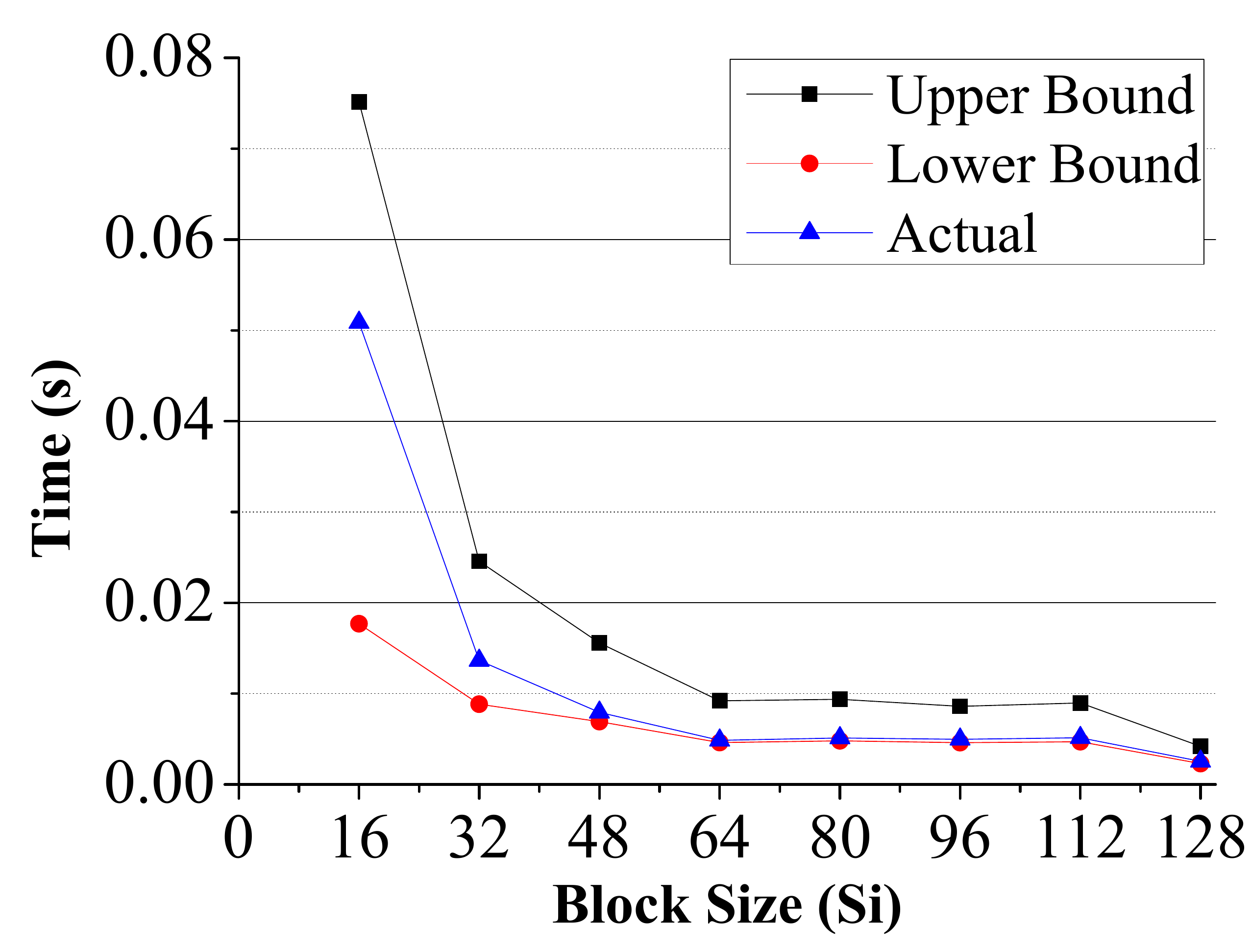}
			\label{128x2}}
		\hfill
		\vspace{-0.5em}
		\subfloat[$N_p=3$]{\includegraphics[width=.245\textwidth]{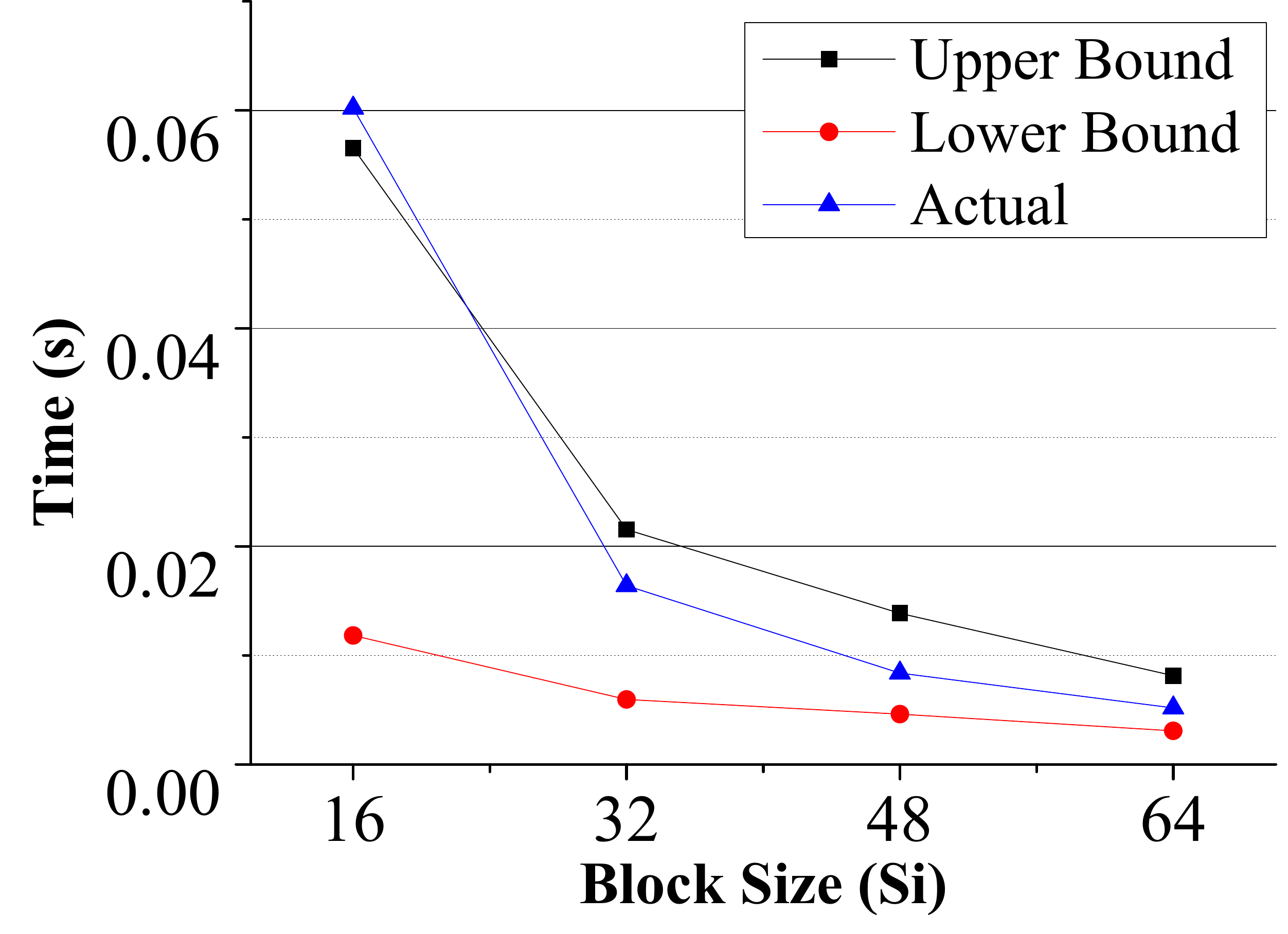}
			\label{64x3}}	
		\subfloat[$N_p=4$]{\includegraphics[width=.245\textwidth]{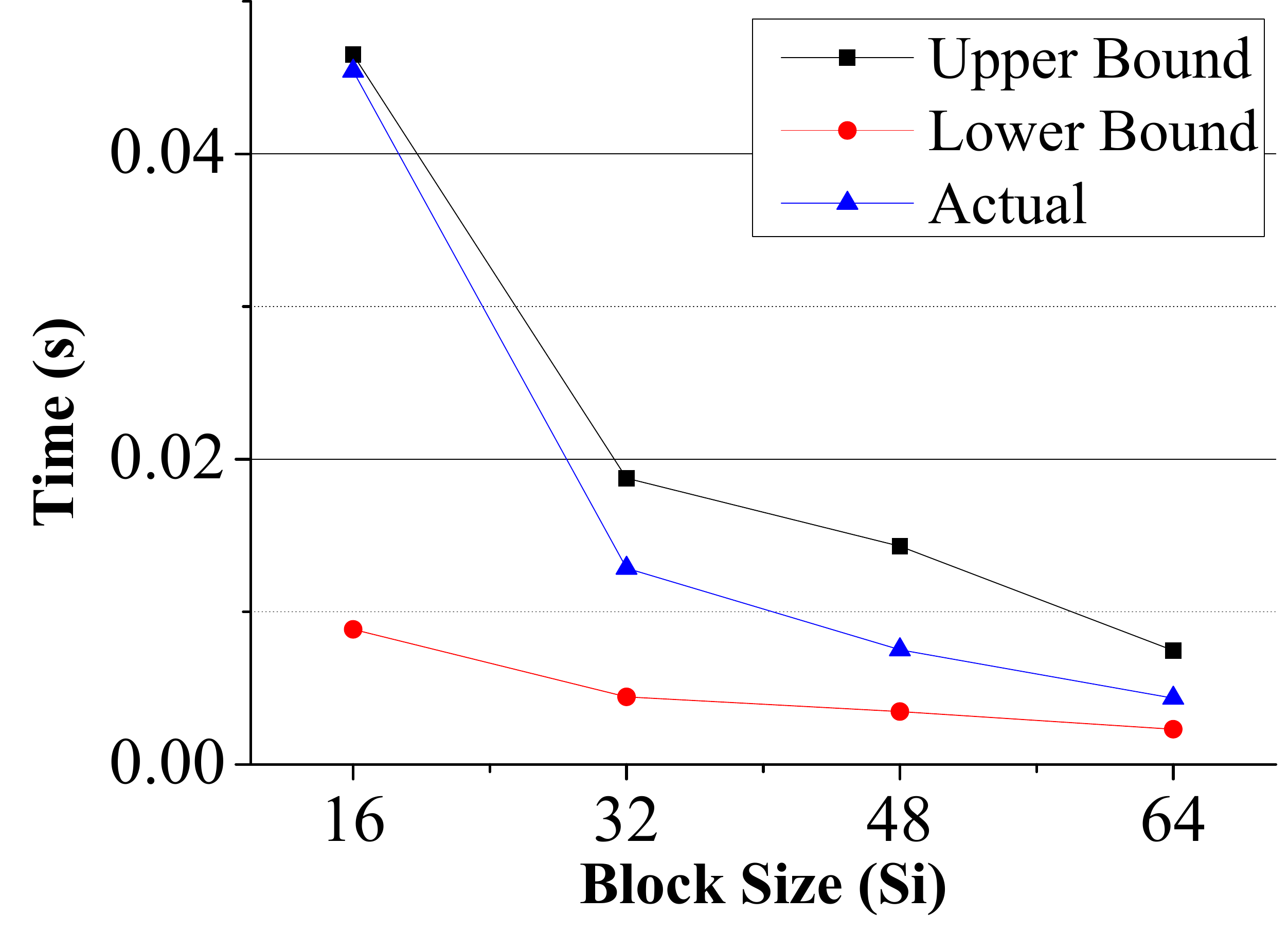}
			\label{64x4}}
		\caption{Comparison of the estimated execution time and actual execution time for conv-2.}
		\label{Analytical Model}
		\vspace{-1.5em}
	\end{figure}
	
	As a case study, we use a real-life CNN model, AlexNet \cite{krizhevsky2012}, to validate our analytical model. Note that AlexNet comprises of five convolutional layers and three fully connected layers, and the computation patterns of these layers can be converted to matrix multiplication \cite{Jason2014}.
   Note that we are mainly focus on determining the optimal design parameters under fixed $P_m$ and $P$, therefore we do not make full use of the resource of the FPGA to pursuit maximum performance. In our experiment, we set $P_m=4$ and $P=64$.
   After post-synthesis, a maximum frequency of 200 MHz ($F_{acc}$) is achieved. 
   Table \ref{resource} summarized the resource utilization of the overall system. It can be seen that the overall resource utilization is below 50\%, which contributes to the high frequency of our accelerator.
   
	We give a detailed comparison between the predicted and actual execution time for conv-2 in Fig.~\ref{Analytical Model}. It can be seen that the predicted lower bound of execution time closely follows the actual measurement when the memory requirement of each PE array is satisfied. However, when the memory bandwidth requirement is unsatisfied, the actual time of each PE array becomes more close to the upper bound of predicted execution time. In addition, it can also be found that using multiple PE arrays does not ensure the optimal performance. For example, the case of $(N_p,S_i)=(1,32)$ achieves lower execution time than the case of $(N_p,S_i)=(2,16)$. The main reason is that both of the cases are memory-bound ($<$ 1.6 GB/s), and the case of $(N_p,S_i)=(1,32)$ can reach higher memory bandwidth (it can be confirmed by Fig.~\ref{Bandwidth}), which contributes to its higher performance.

\begin{table}[!t]
	\caption{Optimal $(N_p,S_i)$ of all layers in AlexNet.}
	\label{table_results}
	\centering
	\begin{tabularx}{\linewidth}{|p{0.835cm}<{\centering}|p{1.7cm}<{\centering}|X|p{0.95cm}<{\centering}|p{0.95cm}<{\centering}|p{0.95cm}<{\centering}|}
		\hline
		\multirow{2}{*}{Layers}&   \multirow{2}{*}{$M*K*N$}    &   Optimal    &  \multicolumn{3}{c|}{Performance (GFLOPS)}    \\ \cline{4-6}
		&                               & $(N_p,S_i)$  & Optimal & $N_p=4$& $N_p=1$ \\ \hline
		Conv-1      & 96*363*3025  &(2,128)& 59.7 &  57.1  &  49.2  \\ \hline
		Conv-2      & 128*1200*729 & (2,128)&87.8 &  70.3  &  61.4  \\ \hline
		Conv-3      & 384*2304*169 & (2, 96)&64.9 &  62.9  &  57.4  \\ \hline
		Conv-4      & 192*1728*169 & (2, 96)&64.1 &  54.8  &  51.2  \\ \hline
		Conv-5      & 128*1728*169 & (2,128)&62.9 &  44.9  &  43.9  \\ \hline
		fc-6        & 128*9216*4096 & (2,128)&100.9 & 79.3 &  70.7  \\ \hline
		fc-7        & 128*4096*4096 & (2,128)&99.3 &  78.1 &  69.5  \\ \hline
		fc-8        & 128*4096*1000 & (2,128)&96.9 &  83.6 &  67.8  \\ \hline			
	\end{tabularx}
	\vspace{-2.3em}
\end{table}
	
   The optimal $<N_p,S_i>$ of all the layers in AlexNet is given in Table~\ref{table_results}. It can be seen that when compared to other extension approaches, i.e extending the number of PEs only ($N_p = 1, P = 256$) and extending the number of PE arrays only ($N_p = 4, P = 64$), our accelerator that implemented with the optimal $<N_p,S_i>$ achieves the highest performance for all layers. In addition, our accelerator achieves 100.9 GFLOPS for fc-6, which demonstrates that our multi-array architecture can achieve high ratio (i.e. up to 98.6\%) of sustained performance to theoretical peak performance (which is denoted by $2 \times F_{acc} \times P_m \times P$ \cite{Dou2005}). 

	\section{Conclusion}
	In this paper, we focus on the architecture extension of the linear array architecture for matrix multiplication on FPGA, by proposing a highly configurable and scalable multi-array architecture. We employ a work-stealing scheme to realize workload balancing among PE arrays. An efficient analytical model is developed to determine the optimal design options for the architecture extension. Experimental results show that our optimal extension of the linear array architecture can reach the highest performance and computation efficiency.
	
	\ifCLASSOPTIONcaptionsoff
	\newpage
	\fi

	\footnotesize
	\bibliographystyle{IEEEtran}
	\bibliography{IEEEabrv,./mybib}
\end{document}